# Seismic Structure of the Southern Rivera Plate and Jalisco Block Subduction Zone

by Diana Núñez, Francisco Javier Núñez-Cornú, Felipe de Jesús Escalona-Alcázar, Diego Córdoba, Jesualdo Yair López Ortiz, Juan Luis Carrillo de la Cruz, and Juan José Dañobeitia

## ABSTRACT

Structural and tectonic features in the Pacific Coast of Mexico generate a high level of seismic activity in the Jalisco block (JB) region, making it one of the most attractive areas of the world for geophysical investigations. The Rivera–North America contact zone has been the object of different tectonic studies in recent years framed within the TsuJal project. To this day, this project is generating numerous crucial geophysical results, which significantly improve our understanding of the region. Our study is focused on the interaction between the south of the JB and Rivera plate (RP), which crosses the Middle America trench. We also cover an offshore–onshore transect of 130 km length between the eastern Rivera fracture zone and La Huerta region, in the Jalisco state. To characterize this region, we interpreted wide-angle seismic, multichannel seismic, and multibeam bathymetry data. The integration of these results, with the local and regional seismicity recorded by the Jalisco Seismic Accelerometric Telemetric Network and by the Mapping the Rivera Subduction Zone experiment, provides new insights into the geometry of the southern RP, which is dipping 12°–14° under the JB in the northeast–southwest direction. Moreover, our results provide new seismic images of the accretionary wedge, the shallow crust, the deep crust, and the upper-mantle structure along this profile.

## INTRODUCTION

The west coast of Mexico is home to a complex convergence of tectonic plates that involves the subduction of both the Rivera (RP) and Cocos (CP) oceanic plates beneath the North American (NOAM) continental plate, along the northern end of the Middle America trench (MAT). The Jalisco subduction zone (JSZ) is located where the RP subducts beneath the Jalisco block (JB), a tectonic unit independent of the NOAM (Luhr *et al.*, 1985; Bourgois *et al.*, 1988; DeMets and Stein, 1990; Allan *et al.*, 1991; Rosas-Elguera *et al.*, 1996; Fig. 1). The convergence direction between the RP and NOAM is almost aligned with the trench-normal direction at the southern end of the JSZ. However, it is parallel to the trench at the northern end of the JSZ (Bandy *et al.*, 2008, 2011; Suárez *et al.*, 2013), this change occurs in the region of the Ipala Canyon, which marks the boundary between the Banderas and the southern Jalisco fore-arc blocks (Urías Espinosa *et al.*, 2016). The onset of plate perpendicular convergence is most likely beneath the Islas Marias (e.g., Bartolomé *et al.*, 2011; Núñez-Cornú *et al.*, 2016). Subduction is the dominant geologic process, which controls the recent seismicity and tectonic deformation within the crust of western Mexico (Bandy *et al.*, 1999). In this area, several large destructive earthquakes occurred in the last century such as the $M_s$ 8.2 and 7.8 in 1932 (Eissler and McNally, 1984; Singh *et al.*, 1985), the $M_w$ 8.0 in 1995 (e.g., Melbourne *et al.*, 1997; Pacheco *et al.*, 1997; Escobedo *et al.*, 1998), and the $M_w$ 7.4 in 2003 (Núñez-Cornú *et al.*, 2004).

The tectonic and structural complexity of this region combined with the seismic hazards around the JB makes the west coast of Mexico very attractive for comprehensive geophysical investigations. Furthermore, another important motivation for this work is the need for a more detailed understanding of the interaction between the RP and the JB along the MAT.

To resolve these questions, Mexican and Spanish scientists investigated this region for RP, CP, and NOAM, in which the interactions among these plates control the tectonics. This investigation was part of the TsuJal project ("Crustal Characterization of the Rivera PlateJalisco Block Boundary and its Implications for Seismic and Tsunami Hazard Assessment") during 2014 and 2016 (Núñez-Cornú *et al.*, 2016). This project was divided into two stages: (1) active geophysical (2014) and (2) passive seismicity investigations (2016).

During the first stage in 2014, a combined onshore–offshore experiment was carried out to acquire marine multichannel seismic (MCS), multibeam bathymetry, potential fields (gravimetry and magnetism), and wide-angle seismic (WAS) data using the RRS James Cook (United Kingdom). In the second stage, from April to September 2016, passive seismicity data were recorded by 10 ocean-bottom seismometers (OBSs), 30 permanent stations from the Jalisco Seismic Accelerometric Telemetric Network (RESAJ; Núñez-Cornú *et al.*, 2018), and 25 portable seismic stations deployed to complement the network. In this article, we present one seismic profile, initially designed only for MCS, but also well recorded by the terrestrial portable seismic stations, and we then combine the resulting seismic profile with the regional seismicity to improve the understanding of the crustal structure and geometry of the contact zone between the RP and the JB.

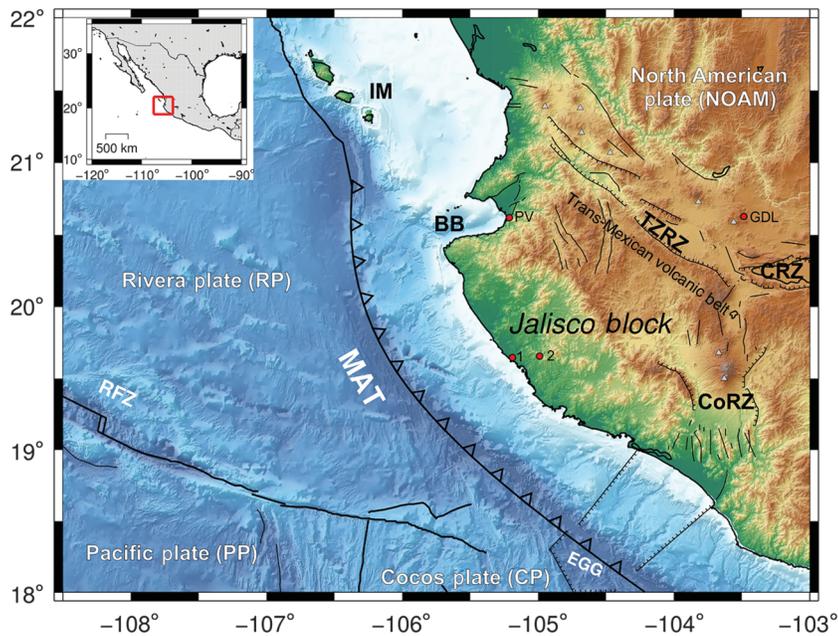

▲ **Figure 1.** Tectonic framework of western Mexico. BB, Bahía de Banderas; CoRZ, Colima rift zone; CRZ, Chapala rift zone; EGG, El Gordo graben; GDL, Guadalajara; IM, Islas Marías; MAT, Middle America trench; PV, Puerto Vallarta; RFZ, Rivera fracture zone; TZRZ, Tepic–Zacoalco rift zone; 1, Pérula; 2, Nacastillo. (Inset) Location map of the study area within the North American continent.

## TECTONIC SETTING

Analysis of the geodynamic interactions presently occurring in the region of the RP is essential to understanding the tectonic complexity of the west coast of Mexico. To this day, the interaction between the oceanic RP and CP remains poorly known due to the absence of clearly associated bathymetric features previously mapped. The boundary between them intersects the Colima rift zone (CoRZ) on land (Eissler and McNally, 1984; Bourgois and Michaud, 1991; Bandy et al., 1995), and the RP separated from the CP from 7 to 10 Ma (Bandy and Hilde, 2000). The southwestern border of the RP corresponds to the recently formed (1.5–3.5 Ma) Rivera transform fault, which presently connects two segments of the East Pacific Rise (Bandy et al., 2011). However, we do not specifically address the interactions of these two plates in this article.

The RP accretes seafloor along its western boundary at the Pacific-Rivera Rise (DeMets and Traylen, 2000). In the northern end of the MAT, the RP subducts beneath the NOAM in the JSZ, in which the convergence direction becomes gradually more oblique toward the northeast (Bandy et al., 2005). From Islas Marías to the south of the JSZ, an uplifted margin along the MAT occurred before the late Miocene. Then during the upper Miocene–lower Pliocene until the Pliocene–Quaternary limit (2.58 Ma), subsidence began (Mercier de Lépinay et al., 1997). Because of that initial period, the RP is considered an independent oceanic plate from the CP.

Eastward to the MAT, the JB is defined as an independent tectonic unit formed by deformation and fragmentation of the overriding continental plate, which is moving slowly to the northwest, relative to the NOAM (Luhr et al., 1985; Bandy and Pardo, 1994). This movement is quantified as 2 mm/yr in a southwestern direction obtained by Global Positioning System (GPS) measurements (Selvans et al., 2011). The geometry of the RP subducting under the JB is still unclear. Some authors suggested a dip angle of 20° or even more (Eissler and McNally, 1984). Other studies reported dip angles of about 12° based on seismicity, WAS, or MCS data (Dañobeitia et al., 1997; Núñez-Cornú and Sánchez Mora, 1999; Bartolomé et al., 2011).

The JB is limited to the north by the extensive structure known as the Tepic–Zacoalco rift zone (TZRZ), which continues to the east with the Chapala rift zone and to the Pacific coast with the CoRZ (Núñez-Cornú et al., 2002). The CoRZ is similar in structure and age to the TZRZ and is defined onshore and offshore by recent seismic activity (Pacheco et al., 2003). The TZRZ consists of several tectonic depressions with extensional and right-lateral motions, which also indicate deep crustal faulting between the JB and the NOAM (Núñez-Cornú et al., 2002). The crustal thickness of the JB along both rift zones has been determined to be about 25 km, and thickening to about 40 km in the central region (Urrutia-Fucugauchi and Flores-Ruiz, 1996). The western JB border is marked by the MAT.

The Rivera fracture zone separates the NOAM, the Tamayo fracture zone, the East Pacific Rise, and the MAT. To the south of the RP is the boundary with the CP that is established between the El Gordo graben and continues at depth east of the CoRZ (Bandy et al., 1995).

## DATA SETS AND PROCESSING METHODS

The region of the RP and the JB has been geophysically sampled by different studies (e.g., Bourgois et al., 1988; Bourgois and Michaud, 1991; Khutorskoy et al., 1994; Bandy et al., 1995, 2005; Michaud et al., 1996; Dañobeitia et al., 1997), but few of them integrated the wealth of onshore and offshore data to obtain a compressive geodynamic model for the western part of Mexico. During the TsuJal project (Núñez-Cornú et al., 2016), multidisciplinary data were recorded both onshore and offshore for the contact region between the JB and the RP. During the active part of the experiment, Mexican and Spanish researchers (TsuJal Working Group) were on board the British research vessel, the RRS James Cook, of the National Environmental Research Council that participated in this project as part of a bilateral scientific exchange agreement between Spain and the United Kingdom (Dañobeitia et al., 2016). This ship was in charge of providing the marine seismic sources, deploying and recovering the OBSs, obtaining the multibeam and high-resolution bathymetry, and obtaining the MCS and potential

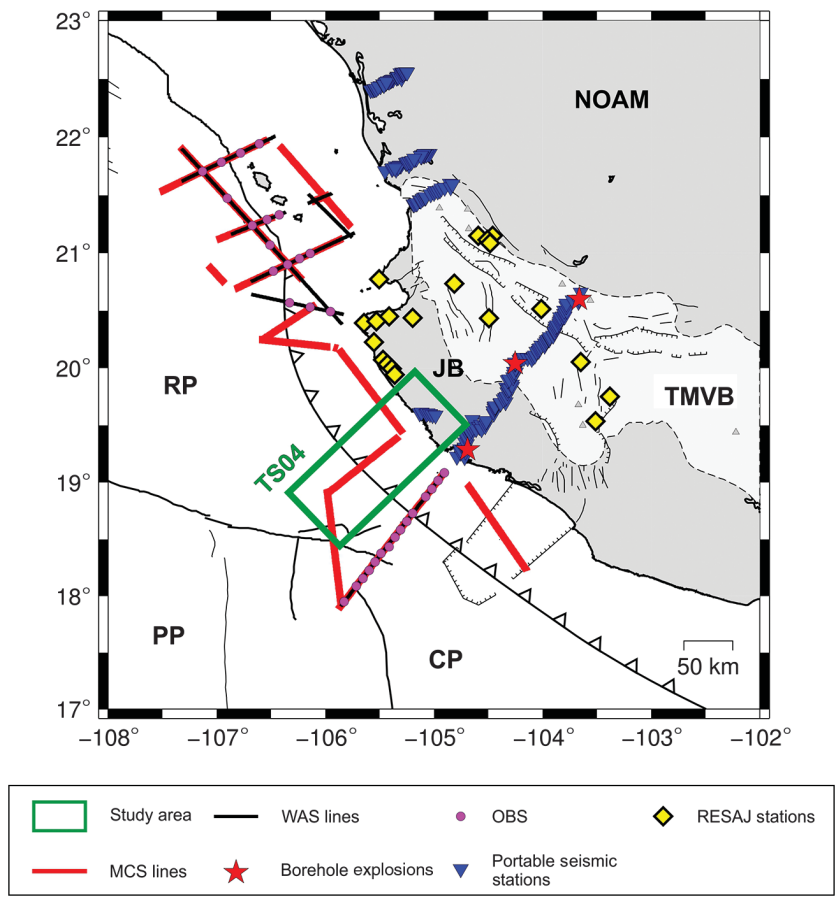

▲ **Figure 2.** TsuJal survey map. Symbols are depicted in the legend and abbreviations in Figure 1. JB, Jalisco block; MCS lines, multichannel seismic lines; OBS, ocean-bottom seismometer; RESAJ, Jalisco Seismic Accelerometric Telemetric Network; TMVB, Trans-Mexican volcanic belt; WAS, wide-angle seismic.

field data during the cruise JC098. This stage was carried out between 17 February and 19 March 2014 (Fig. 2). In this study, we analyze the seismic profile TS04 from the MCS, WAS, and multibeam bathymetry data registered by one streamer, terrestrial seismometers, and two multibeam echo sounders, respectively.

The seismic sources used in this profile aboard the RRS James Cook consisted of 12 BOLT G-GUN 1500LL model guns, the arrangement of which was in four subarrays of three guns each to obtain the maximum energy in the lowest possible frequency range. The total capacity of this array was 5800 in$^3$ with 2000 psi of air pressure, shooting every 50 m and providing 1773 shots in a line 89 km long. The shooting interval was carefully selected to improve the quality of the signal, taking into account the capacity of the air compressors and the maximum data redundancy (Dañobeitia *et al.*, 2016).

## MCS Data
The MCS data were acquired using a streamer with 468 active channels (5850 m) separated 12.5 m apart and deployed at 10 m depth. The common depth point (CDP) distance is 6.25 m providing a CDP nominal fold of 58–59 traces. The MCS data were sampled at 1 ms and recorded in SEG-D format.

In this study, we processed our MCS data with Seismic Unix (Cohen and Stockwell, 2013), and to obtain the best possible seismic image, we followed a traditional processing methodology to increase both vertical and horizontal resolution.

Table 1 summarizes the sequence followed, and the parameters applied to the data. These steps include:

1. prestacked signal calculations (eliminate aliasing, eliminate incorrect traces, and filtering);
2. spherical corrections;
3. predictive deconvolution to improve the temporal resolution;
4. velocity analysis every 100 CDP;
5. correction of normal moveout and stack to increase the signal-to-noise ratio; and
6. phase shift migration with turning rays that increases the horizontal resolution and the collapse diffraction, which relocate the reflectors in time.

## WAS Data
Initially, the configuration of the marine seismic source of the TS04 profile was designed to optimally record only the MCS data; however, seven portable seismic stations aligned perpendicularly to the coast also recorded those sources, providing a WAS transect 130 km in length with a southwest–northeast orientation. These single vertical-component seismic stations (TEXAN 125A geophones, REFTEK, U.S.A.) were installed from Pérula to Nacastillo in the state of Jalisco, Mexico (Fig. 1). Data from one of these stations were not recoverable due to a power failure.

The processing sequence for these data included merging with navigation data, corrections due to instrumental drift, band-pass filtering, as well as travel-time corrections because some stations were located out of the seismic transect (Núñez *et al.*, 2016). Furthermore, we included the bathymetry and topography data (Fig. 3a) to complete the *P*-wave phase interpretation.

The phase correlation consisted of determining the *P*-wave refracted and reflected phases observed at different discontinuities in the crust and the uppermost mantle. The apparent velocities of the refracted waves were determined for the generation of the initial velocity and depth distribution. We identified three refracted phases (one within the sediments [*Ps*], one within the crust [*Pg*] and one within the uppermost mantle [*Pn*]), one crust–mantle boundary reflection (*PmP*), and two reflections in the first layers of the upper mantle $Pm_1$ and $Pm_2$. The *Ps* phase is only observed on the Z000 record section (Fig. 3b,c) between 30 and 37 km offset with an average apparent velocity of 4.5 km/s. The next phase *Pg* is observed in the offset interval between 30 and 64 km, approximately, in all seismic record sections. The apparent velocity is 5.0 km/s. The *Pn* phase is

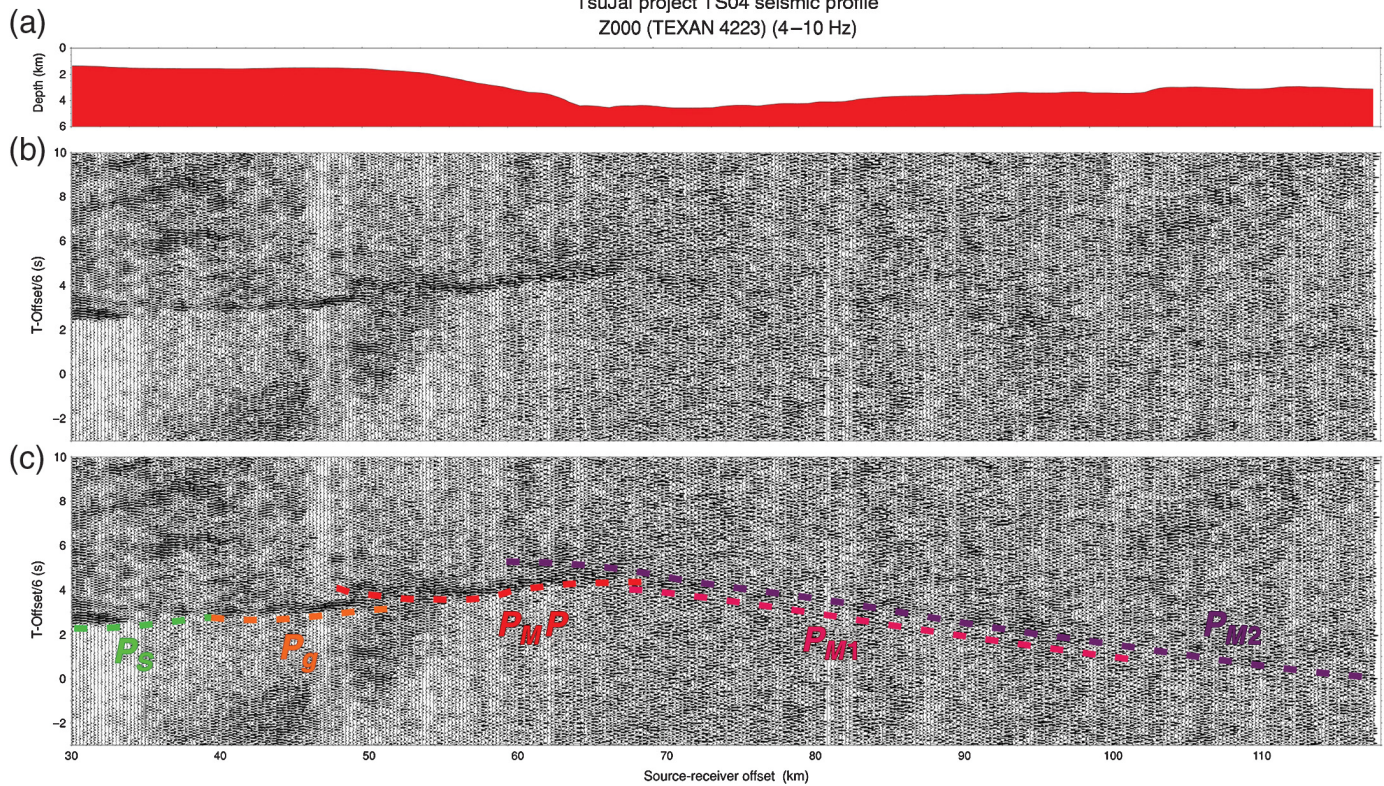

▲ **Figure 3.** (a) Bathymetry along TS04 seismic profile. (b) Seismic record section corresponding to terrestrial seismic station Z000 (vertical component) with a reduced velocity of 6 km/s, 4–10 Hz band-pass filter applied and trace-normalized amplitudes. (c) Z000 seismic record section with interpreted reflected and refracted *P*-wave horizons indicated by different color dashed lines.

correlated in Z002 and Z003 record sections from 66 to 83 and 67 to 95 km, respectively. The mantle-like apparent velocity is more than 10 km/s and could indicate this layer is dipping. All of the record sections display a secondary arrival corresponding to *PmP* reflections, evidencing an abrupt crust–mantle velocity discontinuity. This phase was identified between 47 and 74 km of source–receiver offset distance.

The first secondary arrival $Pm_1$, interpreted to be in the uppermost mantle, is observed at the three stations closest to the coast (Z000 [Fig. 3], Z002, and Z003), whereas the $Pm_2$ is also observed in the Z004 seismic station. The reflected $Pm_1$ phase is correlated between 67 and 98 km offset, whereas it is possible to identify $Pm_2$ up to a maximum offset distance of 125 km.

A total of 441 first arrivals (*Ps*, *Pg*, and *Pn*), *PmP*, and reflections in the upper mantle were manually picked with an average estimated picking error of 82 ms. These WAS data were interpreted using forward modeling, travel-time inversion, and synthetic seismograms with the programs developed by Zelt and Smith (1992) to obtain a final 2D velocity and interface structure model, which best fits the data. Moreover, we used water depth and elevation values from bathymetry and navigation data provided by RRS James Cook and a regional digital elevation model, respectively. We established the origin of the model distance along this transect from a selected point located 117 km southwest of the Z000 terrestrial station.

| Table 1 Processing Flow Applied to a TS04 Multichannel Seismic (MCS) Profile Using Seismic Unix ||
|---|---|
| **Process** | **Parameters** |
| Format change | SEG-D to SU |
| Trace editing | Antialias and bad traces killed (252 and 260) |
| Band-pass filtering | 4–6 to 130–150 Hz |
| Predictive deconvolution | Maximum lag 123 ms, minimum lag 8 ms |
| Geometry specifications | CDP gathering |
| Band-pass filtering | 4–6 to 120–130 Hz |
| Velocity analysis | Semblance spectra, each 100 CDP |
| NMO correction | Velocity model |
| Stack | Velocity model |
| Second predictive deconvolution | Maximum lag 133 ms, minimum lag 5 ms |
| Migration | Phase-shift method |
| CDP, common depth point; NMO, normal moveout. ||

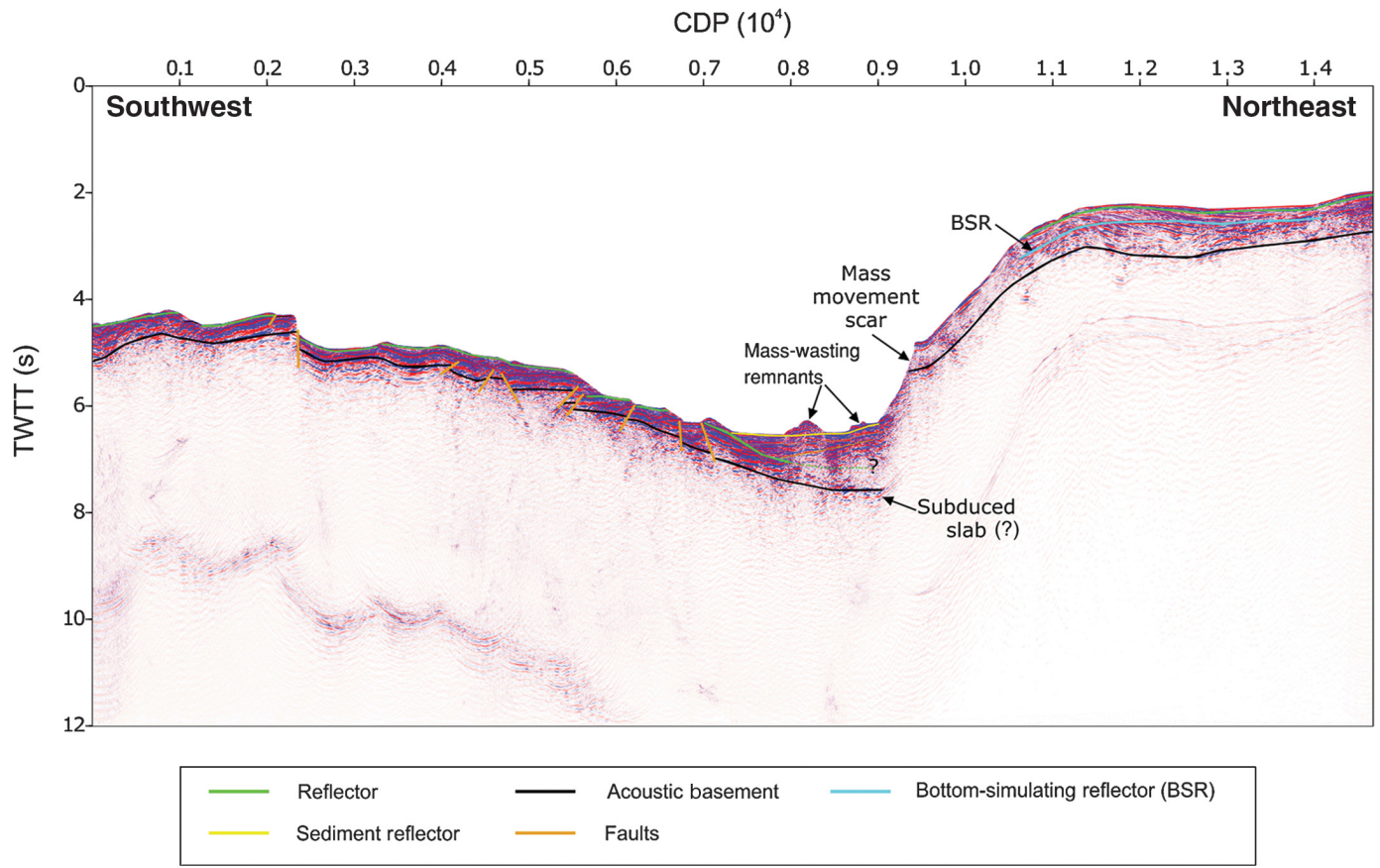

▲ Figure 4. Migrated seismic TS04 line with faults and structures interpreted. CDP, common depth point; TWTT, two-way travel time.

### Multibeam Bathymetry Data

The bathymetry data were collected throughout the TsuJal cruise by means the hull-mounted multibeam Kongsberg Simrad EM120 and EM710 systems. The EM120 was running during the JC098 cruise, whereas the EM710 only ran in the shallowest part of the cruise (<500 m), due to its technical specifications. For our line, we used the data provided by the EM120 echo sounder, the nominal sonar frequency of which is 12 kHz with an angular coverage sector of up to 150° and 191 beams per ping as narrow as 1°. During the cruise, expendable bathy-thermograph probes were used daily to calibrate the sound velocity in the water column.

These data were processed using the CARIS HIPS and SIPS 10.4 (Teledyne) software, a sound speed correction was applied based on speed profiles obtained during the cruise, and a tide correction was applied from data provided by the Centro de Investigación Científica y de Educación Superior de Ensenada. Subsequently, both horizontal and vertical information was combined to produce georeferenced data and to calculate the total propagated uncertainty for each sounding. Finally, regular grid and variable resolution surfaces were generated. Different filters and editors were applied to these surfaces to produce the final bathymetric surface.

## RESULTS

### Bathymetry and MCS data

Figure 4 shows the seismic profile with the most important structures interpreted. In the southwest, we observe normal and inverse faults, which are characteristics of the bending of the RP during the subduction process in the MAT. Moreover, the sedimentary filling of the trench is approximately 1.2 s of two-way travel time (TWTT). The northeast part is characterized by the presence of a bottom-simulating reflector (BSR), and one sedimentary basin was observed between 12,000 and 14,000 CDP with a thickness of 0.25 s of TWTT.

In Figure 5, which is the bathymetric map of the study area, it is possible to directly observe structures of the oceanic crust with no significant sediments in the region of RP until 7000 CDP, near the MAT (Fig. 4). Depths at the MAT range from 3850 to 4900 m in the study area. We identified two landslide areas characterized by a circular shape shown by two arrows in the escarpment region (Fig. 5), and two remnants of mass wasting over the MAT are found in this area, also marked on the figure.

### Wide Angle and Seismicity Data

The final velocity model of profile TS04 (Fig. 6) has seven seismic layers defined primarily by reflections from the first-order

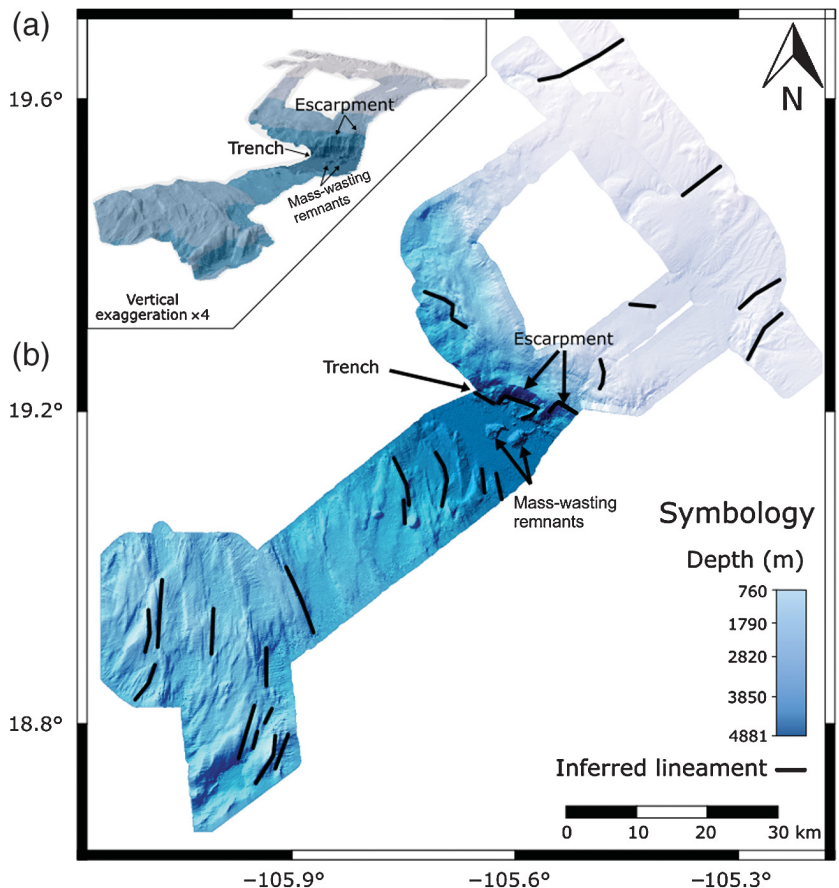

▲ **Figure 5.** Bathymetry of the study area. (a) Bathymetric view of this region with vertical exaggeration ×4. (b) Bathymetric map. Black lines represent the inferred lineaments observed. Two identified mass-wasting remnants and two landslide scarps within the trench are marked. Data obtained with EM120 multibeam echo sounder and processed with CARIS Hips & Sips (v.10.4).

velocity discontinuities. The origin of our model was located at a distance of 117.5 km between the Z000 seismic station and the shot situated farther to the southwest. This model reproduces 99% of observed travel times in the 130 km length of the profile, once adjusted for travel time and controlled by amplitudes through the calculation of synthetic seismograms. For each phase interpreted, the arrival-time fit quality ($\chi_N^2$) was estimated with the following values for $Ps$ (0.9), $Pg$ (2.5), $PmP$ (2.3), $Pn$ (0.6), and $P$ phases observed in the mantle $Pm_1$ (1.5) and $Pm_2$ (2.9). Our final model produces a normalized $\chi^2$ of 2.2, which is not far from the ideal case ($\chi_N^2 = 1$).

The southwestern part of TS04 velocity model is a typical oceanic crust with a thickness of about 9–12 km, including a water layer of a maximum thickness of 4.5 km and a $P$-wave velocity of 1.5 km/s. Up to 45 km of model distance, practically no sedimentary layers are found as also shown in the MCS profile (Fig. 4) and the oceanic crust is 7 km thick with a $P$-wave velocity contrast of 6.8 and 7.9 km/s. In the upper mantle, two seismic layers were defined with velocity increasing with depth and reaching a maximum of 8.5 km/s in the deepest layer.

From 50 to 70 km, two sedimentary layers compose the accretionary prism, with a maximum total thickness of 2.5 km and a $P$-wave velocity range of 2.5–3.0 km/s. From 66 to 79 km, a sedimentary basin was identified with a maximum thickness of 0.6 km and an average $P$-wave velocity of 2.4 km/s, according to the MCS profile. In this region, the RP is subducting under the NOAM with a dip angle of 14°. The Moho discontinuity in this part of the profile has a $P$-wave velocity contrast of 6.6–7.9 km/s.

The northeastern part of the profile, located in the NOAM, is characterized by a seismic layer 1–4 km thick with a $P$-wave velocity range from 4.7 to 5.0 km/s. This layer is underlain by another layer with a thickness ranging between 3 and 7 km and a vertical velocity gradient from 5.3 to 5.9 km/s, in the region close to the accretionary prism, whereas in the continental region of the profile, the velocity gradient ranges from 5.4 to 5.8 km/s. In this region, the velocity in the lower continental crust ranges from 6.8 to 7.0 km/s, and lies above a lower velocity layer of 6.5 km/s. Because of the location of the seismic stations in this study, it was not possible to determine the location of the continental Moho discontinuity in this part of the profile. The maximum total depth reached for the oceanic plate in this study is 40 km.

In this study, we included the seismicity recorded by the RESAJ (Núñez-Cornú et al., 2018) from 2002 to 2013 and by the Mapping the Rivera Subduction Zone (MARS) experiment from 2006 to 2008, manually relocated with HYPO 71 to different depths for better results. The seismicity recorded during the second stage of TsuJal project is still being processed due to the complexity of the OBS records.

In Figure 7, we included the seismicity reported by Núñez-Cornú et al. (2002), Rutz-López and Núñez-Cornú (2004), and Gutiérrez et al. (2015) using data of project MARS, and data recorded by the RESAJ (Núñez-Cornú et al., 2018) from 2010 to 2013 (Núñez-Cornú et al., 2014). The seismicity in the oceanic region is mainly located between 0 and 20 km depth, whereas in the coastal region, the hypocenters reach maximum depths of 40 km. At the northeast end of our profile, we observe an absence of seismicity.

To compare to the WAS and MCS results, we projected the seismicity in a line of 130 km coincident with our 2D velocity model and a width of 20 km (Fig. 8). The comparison between both data sets shows consistency with the structures we propose. Furthermore, we observe that the seismicity data fit the structures of the WAS model, for which we do not have previous data, indicating the slab is subducting in this zone with a dip angle between 12° and 14°, previously reported by Núñez-Cornú et al. (2002).

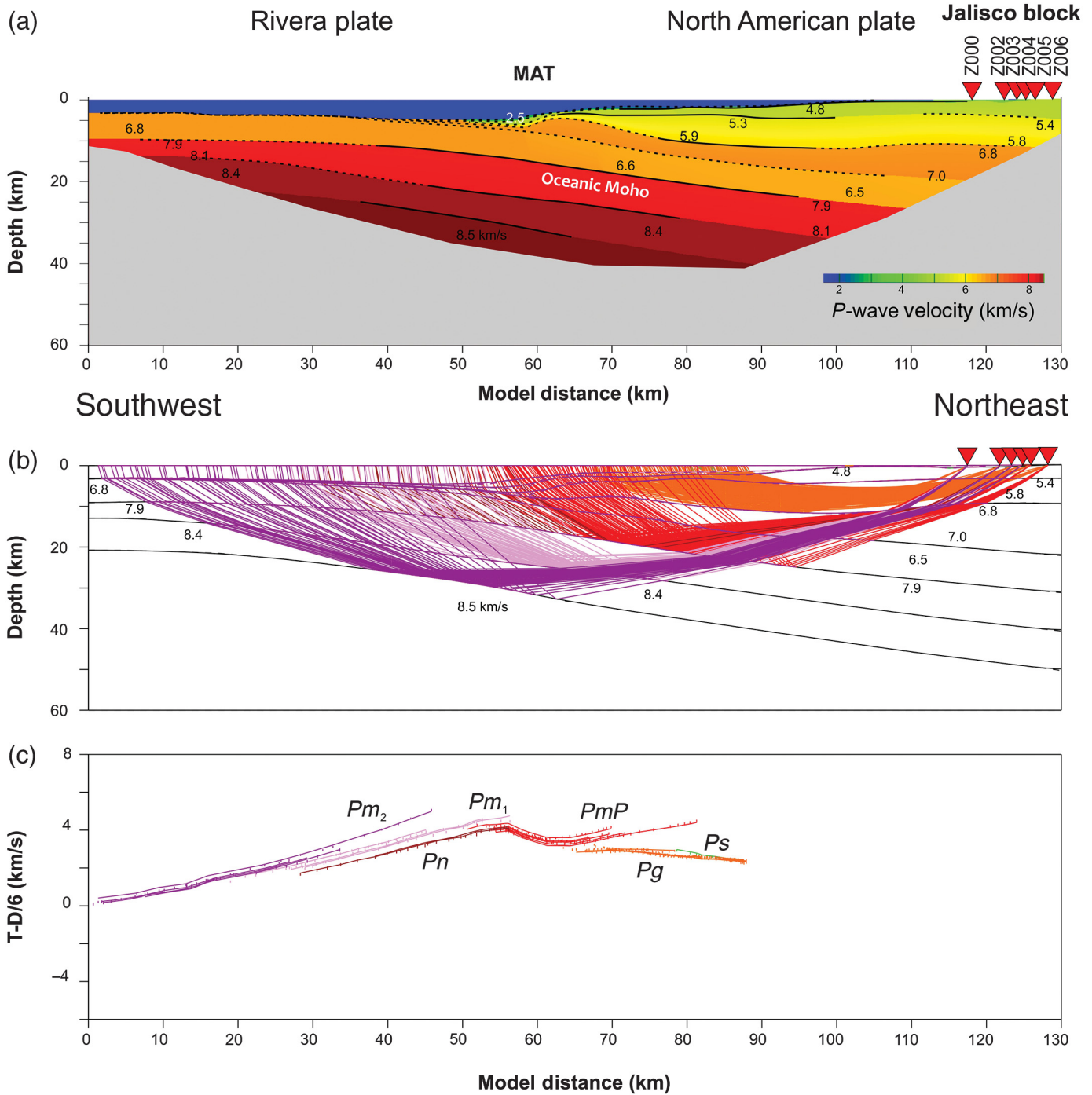

▲ **Figure 6.** (a) TS04 *P*-wave velocity model across the western part of Mexico. Land stations of TsuJal TS04 seismic profile are depicted by red inverted triangles. Vertical and horizontal axes show depth below sea level and model position, respectively. The colored area is the region where ray tracing provides the velocity values. Black lines describe layer boundaries and thick lines mark positions where rays are reflected, showing the well-defined areas. The area not crossed by rays is represented by a gray zone. (b) Ray tracing and velocity model with average velocities in km/s. (c) Comparison between observed (vertical bars) and calculated (lines) travel times. In all panels, distances refer to the velocity model origin.

## DISCUSSION

The analysis of WAS, MCS, bathymetry, and regional seismicity data for the contact between southern RP and JB obtained during the TsuJal experiment confirms a slab's dip angle between 12° and 14°, contradicting the 20° value suggested by Eissler and McNally (1984). Our result is consistent with other seismic studies in the region (Dañobeitia *et al.*, 1997;

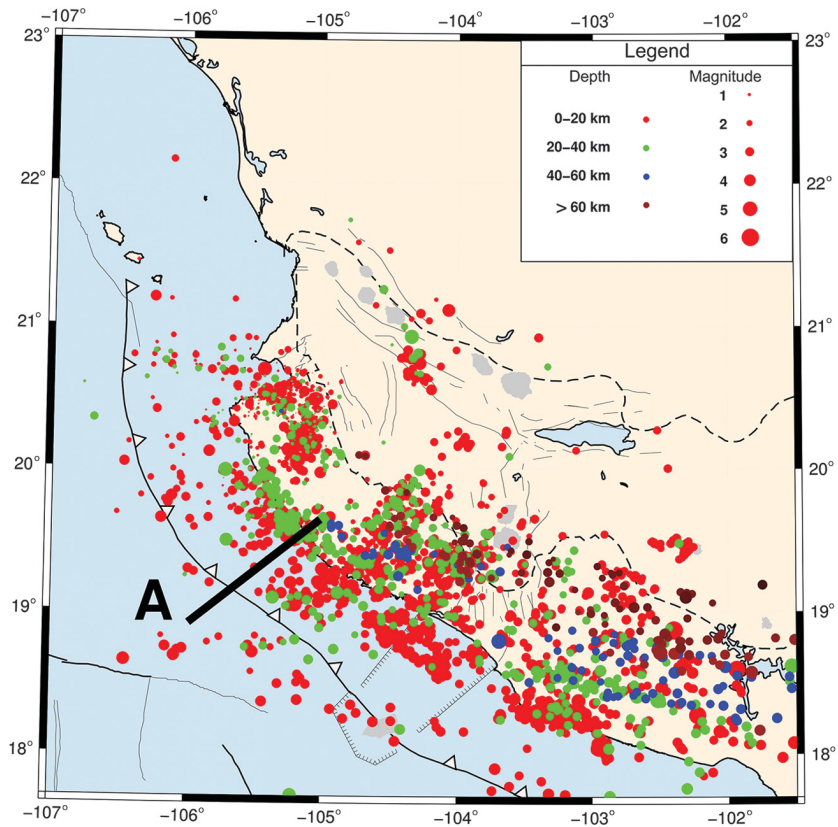

▲ Figure 7. Seismicity map registered by the RESAJ and Mapping the Rivera Subduction Zone (MARS) project (reported by Núñez-Cornú et al., 2002, 2014; Rutz-López and Núñez-Cornú, 2004; Gutiérrez et al., 2015). Profile A is the projection of seismicity along the TS04 seismic profile with a width of 20 km.

Núñez-Cornú and Sánchez Mora, 1999; Núñez-Cornú et al., 2002; Bartolomé et al., 2011). Furthermore, our study reveals new shallow and deep geophysical structures in the crust, which are mainly associated with the subduction process occurring between these plates and the morphological expression of which corresponds to the MAT.

These submarine geological structures and features in the study area are not well described in current literature. In the MCS profile (Fig. 4), we observe the oceanic crust has a bending deflection that forms a palm-tree structure. The small displacements along the normal faults could be due to the 7–22 km thickness of the RP, such that the deformation only occurs in the outermost part of the oceanic crust. We also observe that there is a vertical fault at the point of maximum bending (near CDP 2500; Fig. 4). Because the subduction of the RP is oblique in the area of our study, this structure could be either a lateral fault or part of a scissor fault in which normal and reverse movements take place. In the oceanic crust, none of the imaged reflectors are associated with sedimentary cover; this characteristic was previously described along the MAT by Manea et al. (2003). In the RP region, mainly normal faults are observed, some of which could be active due to the reported shallow seismicity (Fig. 7). Bending of the RP and the shallow seismicity associated with active oceanic crustal faulting have been previously reported (Craig and Copley, 2014; Craig et al., 2014). In the accretionary prism, the morphol-

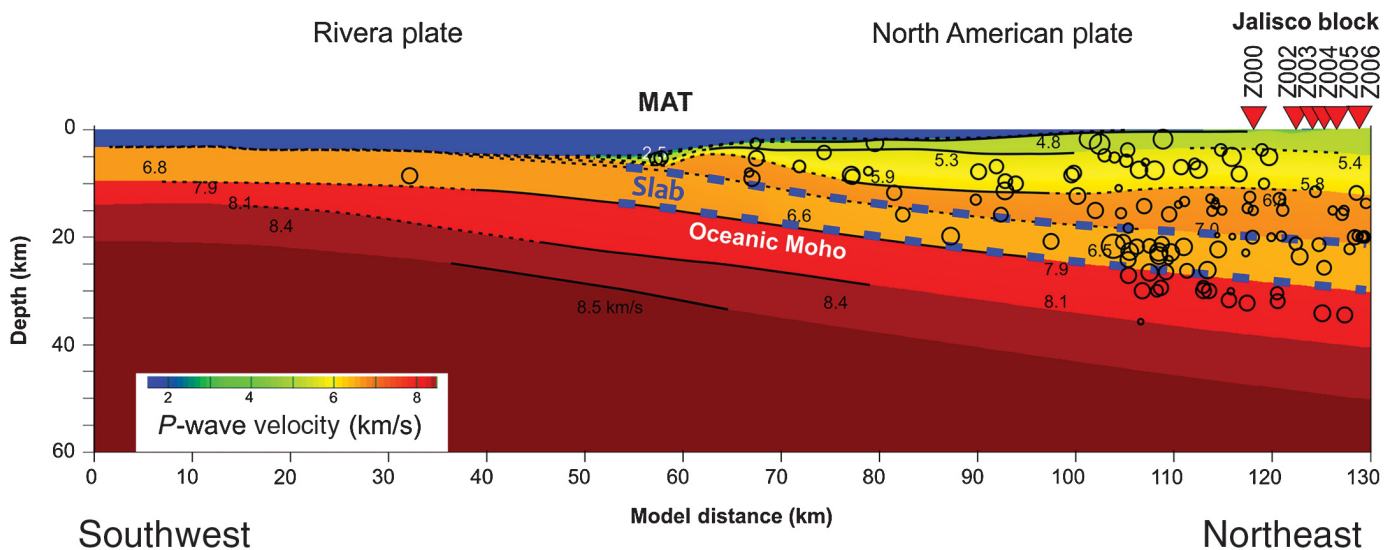

▲ Figure 8. TS04 P-wave velocity model with the seismicity projected along profile A (20 km wide) depicted in Figure 7. Dashed blue lines represent the RP slab subducting beneath the JB.

ogy shows a linear arrangement trending northwest–southeast, parallel to the trench (Fig. 5). These lineaments are identified as reverse faults of the accretionary prism with small relative displacement in the reflectors (Fig. 4). Southeast of the study area, in the El Gordo graben (Fig. 1), the reverse faults and the thin sedimentary cover (polymictic conglomerates and siltstones) were described by Mercier de Lépinay *et al.* (1997). On the continental platform in the JB and over the BSR, the sedimentary cover seems to be thick (<4 km) over an almost flat-lying basement (Figs. 6 and 8) and could be formed by conglomerates and siltstones southwest of the study area (Mercier de Lépinay *et al.*, 1997).

## CONCLUSIONS

This study combines WAS and MCS seismic data, regional seismicity, and bathymetry to characterize the crustal structure in the southern region of the RP along an offshore–onshore seismic profile 130 km in length, perpendicular to the MAT.

Our results show the oceanic crust of the RP is 7–22 km thick with a dip angle of 12°–14°.

The seismicity along the TS04 seismic profile shows a correlation with the structures determined by WAS modeling.

The preliminary study of the MCS profile and the bathymetric map helps identify reverse and normal faults. The normal faults are mostly in the bent oceanic crust, whereas the reverse faults are in the accretionary prism. The sedimentary cover in the oceanic crust is thin, whereas in the accretionary prism it only shows little displacement.

## DATA AND RESOURCES

All geophysical data collected in the TsuJal project and the Red Sísmica y Acelerométrica Telemétrica de Jalisco (Jalisco Seismic Accelerometric Telemetric Network [RESAJ]) are in a database at Research Groups of the Centro de Sismología y Volcanología de Occidente (CA-UDG-276-SisVOc), Universidad de Guadalajara (Mexico), and Estructura y Dinámica de la Tierra, Técnicas de GPS y Estudios Ionosféricos, Universidad Complutense de Madrid (Spain). The data could be used for research joint projects. For information, contact pacornu77@gmail.com or dcordoba@fis.ucm.es. Programs used for processing and modeling data in this study were: (1) multichannel seismic (MCS): Seismic Unix (v.2013); (2) Bathymetry: Caris Hips & Sips 10.4, ArcGIS, QGis; and (3) the wide-angle seismic (WAS): software package developed by Zelt and Smith (1992).

## ACKNOWLEDGMENTS

The authors gratefully acknowledge Wendy McCausland for her valuable comments and observations, including the English grammar revision that notably improved this article. Moreover, the authors would like to thank M. C. Quiriat J. Gutiérrez-Peña for his help with the seismicity maps and Raphael de Plaen for his invaluable assistance with the article. This work could not have been possible without the aid of Rafael Bartolomé and the TsuJal Working Group. The authors are grateful to the guest editors, Editor-in-Chief Zhigang Peng, and anonymous reviewers for their significant comments to improve the scientific content and the English in this article. This research was funded by Consejo Nacional de Ciencia y Tecnología (CONACYT) – FOMIXJAL 2008–96567 (2009) (Mexico); CONACYT – FOMIXJAL 2008–96539 (2009) (Mexico); CONACYT – FOMIXJAL 2010–149245 (2011) (Mexico); CGL (2011-29474-C02-01) DGI Plan Nacional I+D+i (Spain); CONACyT-FOMIXJal (2012-08-189963) (Mexico); NOC Cruise JC098, RRS James Cook (United Kingdom); COIP/COPO/UNAM J-GAP2013 Cruise (BO El Puma); Unidad de Tecnología Marina (Spain); Secretaría de Marina (Mexico) ARM Holzinger; Secretaría de Defensa Nacional (Mexico); Unidad Municipal de Protección Civil y Bomberos (Jalisco State, Mexico); Unidad Municipal de Protección Civil y Bomberos (Puerto Vallarta, Mexico); Unidad Estatal de Protección Civil y Bomberos (Nayarit State, Mexico); Reserva de la Biosfera (Islas Marías) CONANPSEMARNAT; Secretaría de Relaciones Exteriores (Mexico); and Órgano Desconcentrado de Prevención y Readaptación Social de la SEGOP. J. Y. López Ortiz and J. L. Carrillo de la Cruz were financially supported by master fellowships from CONACyT with codes 352189 and 401413, respectively. Some figures were generated using the Generic Mapping Tools (GMT; Wessel and Smith, 1998).

*Diana Núñez*
*Francisco Javier Núñez-Cornú*
*Jesualdo Yair López Ortiz*
*Juan Luis Carrillo de la Cruz*[1]
*Centro de Sismología y Volcanología de Occidente (SisVOc),*
*Centro Universitario de la Costa (CUCosta)*
*Universidad de Guadalajara*
*Avenida Universidad 203*
*Delegacion Ixtapa*
*48280 Puerto Vallarta, Jalisco*
*Mexico*
*diana@sisvoc.mx*
*pacornu77@gmail.com*
*yair.lopez@gmail.com*
*juanluiscc9@gmail.com*

*Felipe de Jesús Escalona-Alcázar*
*Universidad Autónoma de Zacatecas*
*Unidad Académica de Ciencias de la Tierra*
*Calzada de la Universidad 108, Fraccionamiento Progreso*
*98058 Zacatecas, Zacatecas*
*Mexico*
*felscalona@gmail.com*

*Diego Córdoba*
*Universidad Complutense de Madrid*
*Facultad de Ciencias Físicas*
*Plaza de Ciencias, 1, Ciudad Universitaria*
*28040 Madrid, Spain*
*dcordoba@fis.ucm.es*

*Juan José Dañobeitia*
*European Multidisciplinary Seafloor and water column Observatory*
*European Research Infrastructure Consortium*
*Via Giunio Antonio Resti, 63*
*00143 Rome, Italy*
*juanjo.danobeitia@emso-eu.org*

[1] Now at Universidad Nacional Autónoma de México, Instituto de Geofísica, Posgrado en Ciencias de la Tierra, Ciudad Universitaria, 04510 Mexico City, Mexico.